# Experimental demonstration of the suppression of optical phonon splitting in 2D materials by Raman spectroscopy

**Marta De Luca[1‡], Xavier Cartoixà[2‡], David I. Indolese[1], Javier Martín-Sánchez[3,4], Kenji Watanabe[5], Takashi Taniguchi[5], Christian Schönenberger[1,6], Rinaldo Trotta[7], Riccardo Rurali[8*], and Ilaria Zardo[1*]**

1 Departement Physik, Universität Basel, 4056 Basel, Switzerland
2 Departament d'Enginyeria Electrònica, Universitat Autònoma de Barcelona, 08193 Bellaterra, Barcelona, Spain
3 Department of Physics, University of Oviedo, 33007 Oviedo, Spain
4 Center of Research on Nanomaterials and Nanotechnology, CINN (CSIC−University of Oviedo), El Entrego 33940, Spain
5 National Institute for Material Science, 1-1 Namiki, Tsukuba 305-0044, Japan
6 Swiss Nanoscience Institute, University of Basel, 4056 Basel, Switzerland
7 Department of Physics, Sapienza University of Rome, 00185 Rome, Italy
8 Institut de Ciència de Materials de Barcelona (ICMAB–CSIC), Campus de Bellaterra, 08193 Bellaterra, Barcelona, Spain

E-mail:    rrurali@icmab.es
           ilaria.zardo@unibas.ch

**Abstract**

Raman spectroscopy is one of the most extended experimental techniques to investigate thin-layered 2D materials. For a complete understanding and modeling of the Raman spectrum of a novel 2D material, it is often necessary to combine the experimental investigation to density-functional-theory calculations. We provide the experimental proof of the fundamentally different behavior of polar 2D vs 3D systems regarding the effect of the dipole−dipole interactions, which in 2D systems ultimately lead to the absence of optical phonons splitting, otherwise present in 3D materials. We demonstrate that non-analytical corrections (NACs) should not be applied to properly model the Raman spectra of few-layered 2D materials, such as $WSe_2$ and h-BN, corroborating recent theoretical predictions [*Nano Lett.* **2017**, 17 (6), 3758-3763]. Our findings are supported by measurements performed on tilted samples that allow increasing the component of photon momenta in the plane of the flake, thus unambiguously setting the direction of an eventual NAC. We also investigate the influence of the parity of the number of layers and of the type of layer-by-layer stacking on the effect of NACs on the Raman spectra.

## 1. Introduction

Inelastic light scattering is one of the most widespread non-destructive characterization techniques used in materials science. In the increasingly important field of van der Waals two dimensional (2D) materials, such as graphene, hexagonal boron nitride (h-BN), and transition metal dichalcogenides (TMDs), it plays an important role in the determination of the thickness of few-layer flakes. Besides this common use, Raman spectroscopy is also employed for carrying out a thorough investigation of the flakes' lattice dynamics to gather information on exciton-phonon interaction [1], interlayer charge transfer dynamics of van der Waals heterostructures for optoelectronic applications [2], electronic properties [3], and thermal transport properties of the flakes [4]. For all these reasons, it is crucial to achieve a complete and reliable interpretation of the Raman spectra of these materials.

*Ab initio* calculations based on density functional perturbation theory (DFPT) are a valuable tool to assist in the interpretation of the experimental Raman studies, and have demonstrated to be able to predict the vibrational properties of solids with great accuracy [5][6][7]. In polar materials a good agreement with the experiments is obtained only when the long-range dipole−dipole interactions, deriving from the dynamical electric dipoles induced by longitudinal optical (LO) phonon modes, are properly accounted for. This is done by means of the so-called non-analytical corrections (NACs), which modify the dispersion relation next to the Brillouin zone center and lift the degeneracy between the transverse optical (TO) modes and the LO [5][8][9].

The use of NACs in polar 2D materials is not straightforward. For a start, in a DFPT calculation of the Raman spectrum the direction along which NACs should be applied is obtained from the directions of the incident and scattered light, because the difference in photon momenta must equal the momentum of the generated phonon; as a matter of fact, only selecting the proper NAC direction yields results in agreement with the experiments in the different scattering geometries [10]. In backscattering geometry, the direction of the incident and scattered light is usually the one perpendicular to the plane of the layered material, and this poses a problem in the calculations because, since the system is not periodic along that direction, **q** is not a good quantum number and NACs are ill defined. Second —and most importantly— the long-range behavior of a 2D vs 3D dipole lattice is different, and extending the formalism developed for the latter to the two-dimensional case is questionable. This issue has been recently addressed in detail by Sohier and coworkers [11]. They provided theoretical arguments showing that, when treating properly dipole-dipole interactions in 2D (i) the TO-LO splitting vanishes at Γ and (ii) at small wavevectors the LO phonon branch increases linearly with momentum. These findings have far reaching consequences for the understanding of lattice dynamics of 2D layered materials and for the predictions of their dispersion relations. However, with regard to a DFPT calculation of the first-order Raman intensities, they reduce to the simple prescription that NACs, which would otherwise force a TO-LO splitting, should not be used.

In this work we provide experimental evidence based on Raman spectroscopy of few-layer $WSe_2$ and h-BN that TO-LO splitting is indeed absent in polar 2D materials, and that an agreement with theoretical predictions within *ab initio* DFPT can only be attained if NACs are not included in the calculations. We also perform measurements on tilted substrates, with the purpose of increasing the fraction of photon momenta that is transferred to the plane of the material, thus removing the ambiguity in the choice of the NAC-direction, should they be needed. Finally, we discuss the effect of the parity of the number of layers and the role of inversion symmetry on the effect of NACs on the computed Raman spectra.

## 2. Results

### 2.1 Samples

This work focuses on the Raman investigation of thin-layers of $WSe_2$ and h-BN. In particular, we have measured flakes containing monolayer (1L), bilayer (2L), trilayer (3L), and quadrilayer (4L) of $WSe_2$ and flakes containing 1L and 3L of h-BN.

To obtain the $WSe_2$ flakes, the bulk crystal was first mechanically exfoliated on a polydimethylsiloxane (PDMS) stamp (see also the Methods section). Therein, the exfoliated flakes were localized in an optical microscope by transmission measurements (the PDMS stamp being transparent). The desired flakes were deterministically transferred onto a $SiO_2$ substrate by using a micro-manipulator. Optical images and atomic force microscopy (AFM) analysis can be found in ref [12]. The h-BN flakes were obtained by exfoliating bulk BN crystals using the adhesive tape method and were transferred from the tape to a silicon wafer with an oxide thickness of 85 nm. The AFM characterization of the h-BN flake is shown in Supplementary Figure 1 in section 1 of the Supporting Information (SI).

*2.2 Polarization dependent Raman scattering*

Polarization-resolved Raman scattering experiments and DFPT calculations were performed in backscattering geometry (see also methods section). The incident (scattered) photon wavevector is antiparallel (parallel) to the z axis, which is perpendicular to the plane of the flake (xy) when the flake is not tilted. Spectra were collected (and calculated) by selecting scattered light polarized either parallel or perpendicular to the polarization direction (x) of incident light, namely in the xx or xy scattering geometries, respectively. Due to energy and momentum conservation laws, first-order Raman scattering can typically provide access only to phonons at the $\Gamma$ point.

*2.2.1 $WSe_2$*

Let us start our Raman-DFPT investigation with $WSe_2$. Based on our calculations, as well as on previous experimental observations ([12] and references therein), we expect four peaks in a Raman spectrum of $WSe_2$ flakes (~176, ~248, ~250, ~310 cm$_{-1}$). Below we discuss the modes at ~248 and ~250 cm$_{-1}$. The observation and interpretation of the modes at ~176 and 310 cm$_{-1}$ is instead provided in Ref. [12] and is not directly relevant here.

In Figure 1 we show the Raman spectra magnified in the region of the $A'_1$ (or $A_{11g}$) and $E'$ (or $E_{1g}$) modes at ~250 cm$_{-1}$. For the sake of clarity, in the following discussion we will only use the symmetries of the 1L to label them, i.e. $A'_1$ and $E'$. The first column refers to experimental spectra (taken with the excitation wavelength $\lambda_{exc}$=633 nm), while the second and third columns display the theoretical spectra calculated without and with NACs applied for **q** in the plane of the flake. Let us first focus on the theoretical spectra without NACs, which nicely reproduce the experimental spectra. In the experimental and in the theoretical spectra without NACs, the low-frequency mode ($E'$) downshifts with increasing sample thickness, while the high-frequency mode ($A'_1$) upshifts. Supplementary Table 1 in the SI2 shows the frequencies of all the phonon modes (Raman active or not) of the 1L, 2L, 3L, and 4L calculated by DFPT. Clearly, the A'$_1$ and E' modes are almost degenerate in the 1L (the two E' modes correspond to the degenerate TO$_2$/LO$_2$ modes at 250.76 cm$_{-1}$ and the A'$_1$ mode to the single ZO$_2$ mode at 251.31 cm$_{-1}$), for which basically a single peak at ~250 cm$_{-1}$ is observed in xx configuration (made by a convolution of the TO$_2$ and ZO$_2$ mode). In xy geometry only the LO$_2$ mode is allowed and contributes to the Raman signal. In thicker layers, there are several, and very close in frequency, phonon modes in the 250 cm$_{-1}$ range that are Raman allowed in our geometries. Due to the experimental lineshape broadening, some of these modes are merged and become indistinguishable. Therefore, in Supplementary Figure 2 in SI3 we display for all the WSe$_2$ samples the calculated Raman spectra generated by using Lorentzian curves with a full width at half maximum (FWHM) much narrower (0.1 cm$_{-1}$) than the experimental one (1.5-2.0 cm$_{-1}$), such that the different contributions are well visible. By correlating Supplementary Table 1 and Supplementary Figure 2, it is clear that in the 2L sample the low-frequency component (visible with the same intensity in xx and xy in the theory spectra) is assigned to the two degenerate modes at 249.57 cm$_{-1}$ (whose symmetry is E$_{1g}$), and the high-frequency component (visible only in xx) to a single A$_{11g}$ mode with frequency 251.79 cm$_{-1}$. In the 3L sample, there are two low-frequency components in the theory spectra in both xx and xy geometries, which results in a splitting of the low-frequency component of the calculated spectra in Figure 1 and in a broadening of the corresponding experimental peak. Based on the Supplementary Table 1 and Figure S2, we attribute the low-frequency component of the xx and xy 3L spectra to a convolution, weighted by their calculated intensity, of six E'$_1$ modes: two modes at 248.57cm$_{-1}$, two modes at 249.60 cm$_{-1}$, and two modes at 249.61cm$_{-1}$. The high frequency component of the xx spectra is attributed to a convolution of two single A'$_1$ modes, at frequency 250.08 and 252.01 cm$_{-1}$, the latter being more intense than the first, which results in the appearance of a single peak. Similarly, the 4L sample exhibits four Raman active modes of symmetry E$_{1g}$ (two modes at 248.49 cm$_{-1}$ and two modes at 249.59 cm$_{-1}$) visible both in xx and xy scattering configurations and two Raman active modes of symmetry A$_{11g}$ (one at 250.62 cm$_{-1}$ and one 252.12 cm$_{-1}$, the latter being much more intense than the first) visible only in the xx scattering configuration.

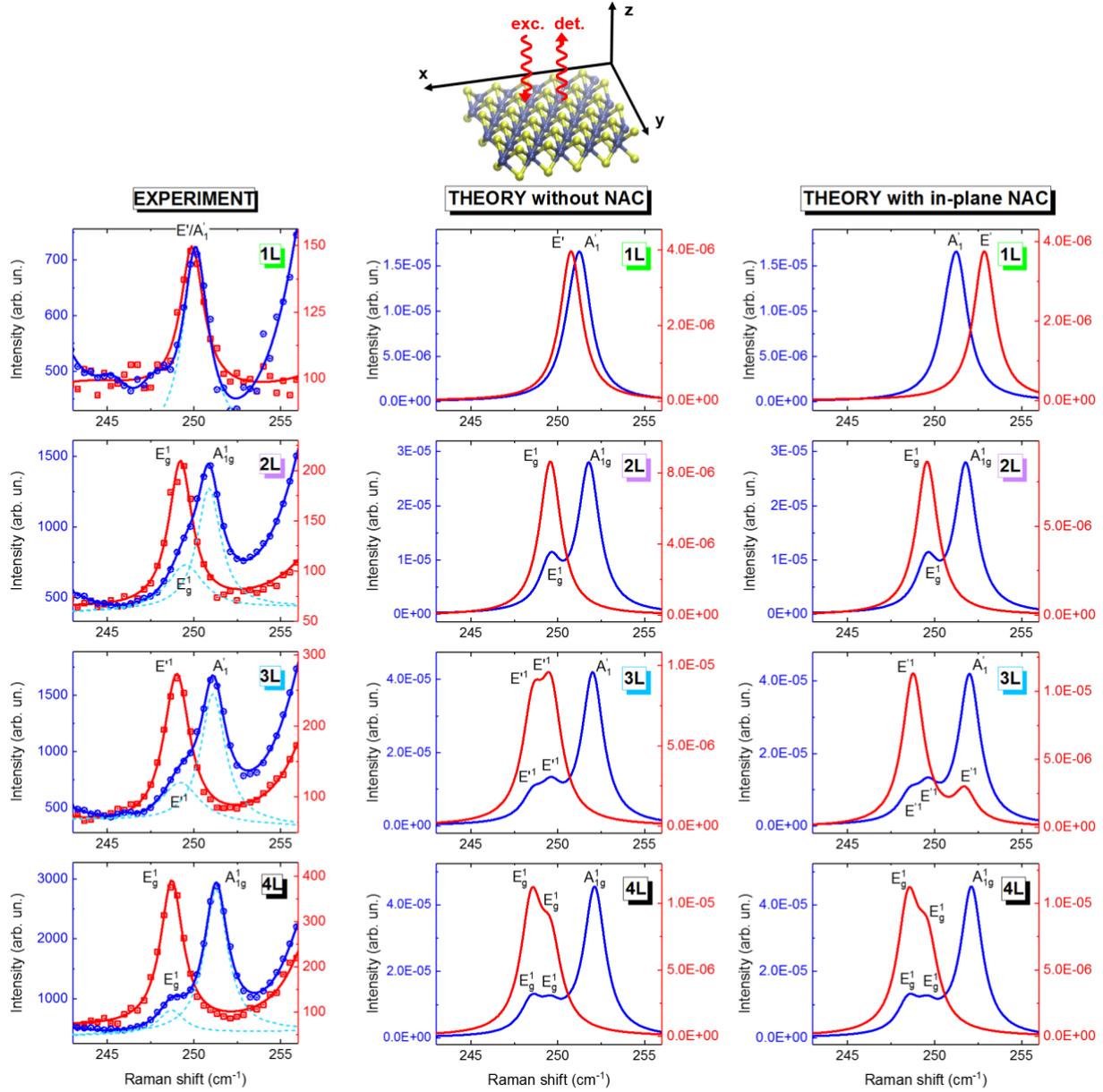

**Figure 1.** Experimental (first column) and calculated (second and third columns) Raman spectra of the 1L, 2L, 3L, and 4L $WSe_2$ samples in the xx (blue lines and circles) and xy (red lines and squares) scattering geometries magnified in the region of the first-order phonon modes at ~ 250 cm$^{-1}$. The theoretical spectra are calculated without NACs (second column) and with NACs in the flakes' plane (third column) and a FWHM of 1.5 cm$^{-1}$ has been used for the Lorentzian peaks (see methods section). Experimental spectra were acquired with the 633 nm excitation wavelength. Solid lines in the experimental spectra are fits to the data (in the spectra taken in the xx configuration, in which two peaks are present in the 2L, 3L, and 4L, we also display the single Lorentzian components by dashed lines). For each couple of spectra (xx and xy), intensity scales (shown on the left and right, respectively) were chosen in order to approximately align the background signal as well as the maxima, to facilitate the comparison between the two polarized spectra and also with the theory. The top inset is a sketch of the experimental and theoretical scattering geometry (we display a monolayer for representative purposes).

In order to compare our experimental results with theoretical spectra obtained with NACs, we must first determine the proper NACs direction. Given that most Raman experiments are carried out in a backscattering geometry at perpendicular incidence with respect to the flake, it would be tempting to take the z direction as the one along which the NACs should be computed. Indeed, if one takes this approach, no splitting is obtained from the calculations, in agreement with experiments. This accidental agreement might be the reason why the absence of NAC-induced splitting has not previously received more attention. However, there is an obvious problem with taking the NACs axis as z. While it is true that backscattered photons have had their change of **k** mainly along z, this momentum kick is not provided by a phonon with a well-defined $q_z$. Rather, the strong confinement of the vibrational mode in the $z$ axis can be thought of as containing a large superposition of $q_z$'s. In other words, the non-periodic character of the layer along z renders $q_z$ an undefined quantity, forbidding any approach to Γ along the z-axis.

On the other hand, given that Raman-scattered photons are measured, phonons must have been created in the flake and, since the flake is planar, these phonons must be characterized by some in-plane **q**$_{//}$, which readily indicates that the correct NAC direction to consider for comparison with the experiment is in-plane. The specific direction will depend on the specific **q**$_{//}$ of the phonon generated and we know that phonons must be generated isotropically in the plane. However, the macroscopic equivalence of the x and y axes means that all the in-plane directions will yield the same amount of splitting, so we finally settle for NACs along the x-axis as representative of the physically meaningful NACs approach direction in our experimental geometry.

With these considerations in mind we now turn our attention to the theoretical spectra obtained with in-plane NACs, in the common formalism developed for bulk 3D materials, and compare them with the experimental results. Basically, the NACs introduce a TO-LO splitting, which impacts the spectra in the xy geometry as the LO mode is allowed only in that geometry. In the 1L in xx scattering configuration the $TO_2$ mode is at 250.76 cm$_{-1}$ and the $ZO_2$ mode at 251.31 cm$_{-1}$, as in the case without NACs (given the finite FWHM of the modes, the $TO_2$ and the $ZO_2$ modes also here are indistinguishable in the calculated spectrum).

In xy, instead, only the $LO_2$ mode is visible, but it shifts at 252.86 cm$^{-1}$, which is in disagreement with the experimental results (those in Fig. 1 as well as all the literature, *e.g.*, in refs. [13][14][15][16]), where the $LO_2$ mode does not shift. In the case of 2L including or not the NACs does not make a difference in the Raman spectra, regardless of the scattering geometry considered. The reason is that in this case the modes whose frequencies are altered by the NACs are forbidden in backscattering geometry and thus the spectra do not change, as schematically illustrated in Figure 2. This is a general feature of systems with inversion symmetry such as even-layered $WSe_2$, as we have indeed verified in the 4L. The fact that a good agreement with the experiments can only be achieved in odd-layer flakes without including the NACs corroborates the predictions of ref. [11]. We do not discuss phonon modes at 176 and 310 cm$^{-1}$ (see ref. [12] for more details), as the calculated Raman spectra do not change upon application of NACs. This occurs regardless of the parity of the layers, as exposed in Supplementary Figure 3 in SI4.

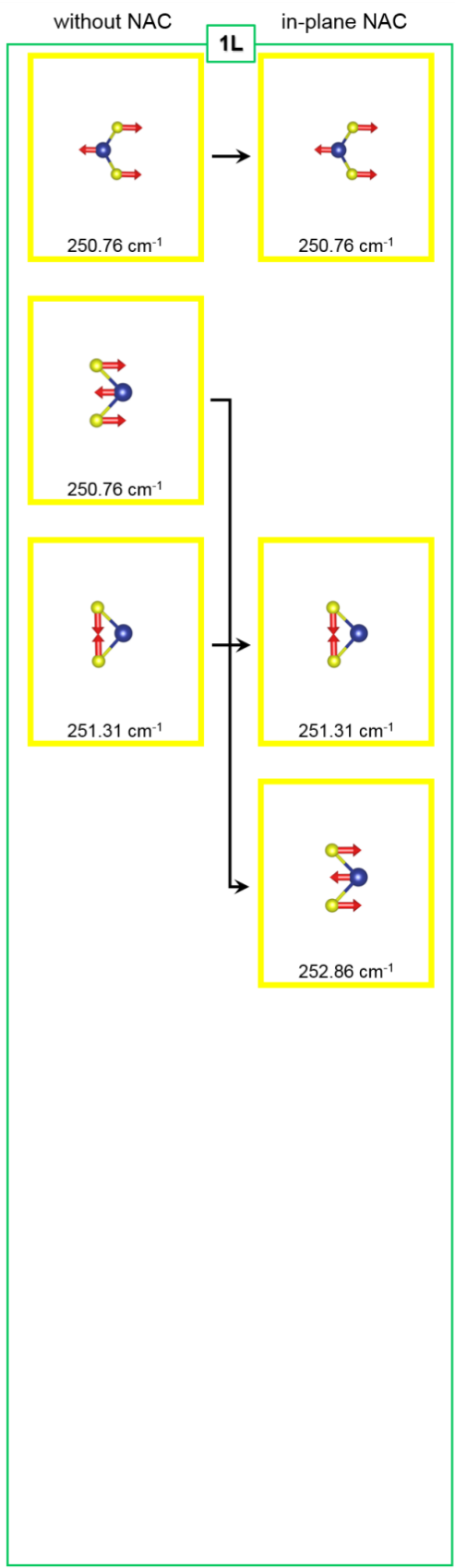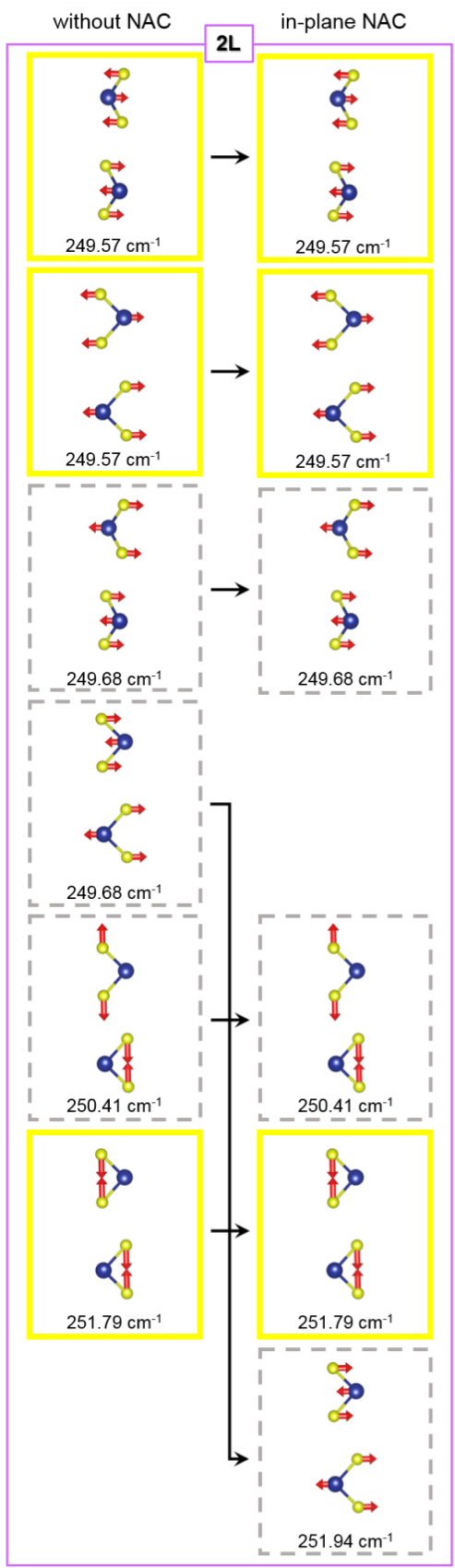

**Figure 2.** Sketch of the atomic displacements of all the phonon modes around 250 cm$^{-1}$ for the 1L (left panel) and 2L (right panel) WSe$_2$ samples, without NACs (left column) and with NACs (right column) in the flakes' plane, as indicated. Displacements contained in yellow solid line frames correspond to Raman-allowed modes in our scattering geometries (xx, xy, or both), while displacements contained in grey dashed frames correspond to modes that are either Raman-inactive or forbidden in our geometries. Big blue spheres represent W atoms and small yellow spheres Se atoms. Black arrows are a guide to follow the effect of the application of in-plane NACs on each mode.

In the following, we discuss measurements performed by tilting the flake in the xz plane, such that the momenta of the exciting and detected photons are still directed along the z axis (backscattering geometry), but the flakes are lying on a plane that forms an angle of 23° with the xy plane (see the sketch on the top part of Figure 3). The scattering geometries, xx and xy, are defined as in the planar case. In the tilted configurations, the momentum of the exciting (and detected) photon that is transferred in the plane of the material is finite making less ambiguous the choice of NACs in the flakes' plane, and the comparison between the experiment and the two types of calculations more direct. We have performed this experiment on the 1L, since, as shown in Figure 1, it is the sample in which the two theoretical approaches differ and thus the experimental results can unambiguously establish which one is correct. We stress that the tilt angle and the relative orientation between the polarization vectors and the flake plane are taken into account also in the calculations, regardless of the application of the NACs. The experimental results are shown in the first row of Figure 3, the calculation without NACs in the second row, and the calculations with NACs computed along the in-plane component of the incident/backscattered photon **k**-vector in the third row. Clearly, the experimental spectra collected on tilted flakes are very similar to those collected on the planar flake and displayed in Figure 1. Moreover, similarly to what happens in the theoretical spectra in Figure 1, the application of in-plane NACs upshifts the LO$_2$ mode in the xy geometry, which is in disagreement with the experimental results. The lack of the appearance of an E' mode in the xy configuration at frequency higher than that of the A'$_1$ mode further proves that the TO-LO splitting in 2D materials is suppressed and unambiguously establishes that the NACs in the formalisms developed for bulk 3D materials should not be used. We point out that the computed spectra in Figure 3 relative to the tilted sample (without and with NACs) are very similar to those in Figure 1 relative to the planar sample, with the exception of a decrease in the intensity. This effect is not a NAC-related effect: it simply derives from the different intensity that phonon modes have in the different scattering geometries (*i.e.*, to selection rules, as explained in the methods section). The change in the intensity is visible also in our experimental measurements. However, its quantitative assessment is challenging, since the intensity reduction expected for the considered tilt angles is sufficiently small that also the effects of the finite numerical aperture of our objective would need to be accounted for [10][17].

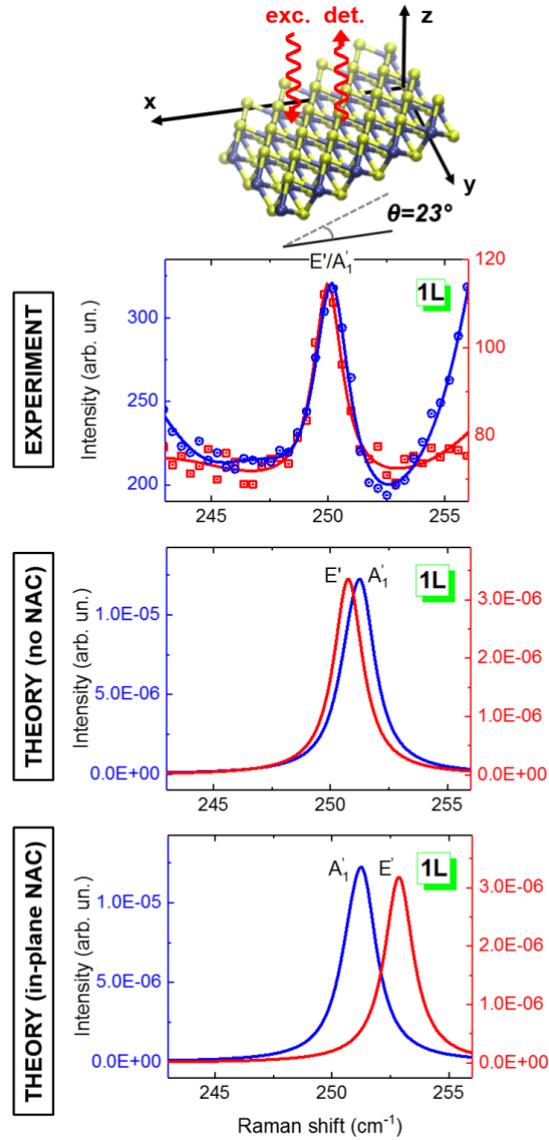

**Figure 3.** Experimental (first row) and calculated (second and third rows) Raman spectra of the 1L flake of $WSe_2$ in the xx (blue lines and circles) and xy (red lines and squares) scattering geometries in the ~ 250 cm$^{-1}$ range. The theoretical spectra are calculated without NACs (second row) and with NACs applied in the flake's plane (third row). Spectra were measured/calculated for a flake lying on a plane forming 23° with the x axis, as indicated in the sketch on the top. Experimental spectra were acquired with the 633 nm excitation wavelength. Solid lines in the experimental spectra are fits to the data. In the experimental spectra, intensity scales were chosen in order to approximately align the background signal as well as the maxima. The intensity scales in the theory spectra calculated without NACs spectra were arbitrarily chosen (see left and right scales) to align the maxima in xx and xy, similar to what

was done in Figure 1, and the same scales were used for the spectra calculated with in-plane NACs. Notice the decrease in intensity compared to the pertinent theory spectra in Figure 1.

*2.2.2 h-BN*

To further corroborate our results concerning the suppression of the TO-LO splitting, we have also studied h-BN, an atomically-thin 2D materials that features some important differences with WSe$_2$. In particular, as we discuss below, the presence of NACs would induce a much larger TO-LO splitting and, consequently, it is much easier to rule out its existence. The analysis is further simplified by the fact that the flexural ZO mode is well separated from the other optical modes, which, because of the very strong B-N covalent bonds, lie at very high frequency [18][19]. The in-plane vibrational modes, the TO and LO modes, are degenerate. The experimental frequency of the TO/LO mode is ∼ 1370 cm$_{-1}$ in 1L h-BN and decreases with increasing layer thickness [19].

Figure 4 shows the experimental results (first row) obtained with a 1L h-BN flake at two different tilt angles (0° and 23°), both in the xx and xy scattering geometry. The calculated Raman spectra for the 1L with FWHM= 9 cm$_{-1}$ without NACs are displayed in the second row, and with NACs along the in-plane component of the photon **k**-vector in the third row. The calculated Raman spectra for planar 1L, 2L, and 3L samples with narrow FWHM (with and without NACs) are displayed in Supplementary Figure 4 in SI5, where all the contributions to the Raman signal are clearly distinguishable. Figure 4 shows that for the 1L the comparison with the computed Raman spectra is very good, provided that NACs are not included (although the frequency of the modes is somewhat overestimated in the theory, as discussed in the methods section). Instead, including the NACs in the calculation introduces qualitatively different features in the predicted Raman spectrum, such as the lifting of the degeneracy between TO and LO, with the latter shifting to ∼1504 cm$_{-1}$ in the xy scattering geometry, which is in disagreement with the experimental observations. We note that the huge TO-LO splitting predicted by the NACs (∼ 115 cm$_{-1}$) would make the experimental observation of the LO unambiguous, as its absence cannot be attributed to lack of resolution or to broadening of the phonon linewidths. Moreover, from the experimental point of view, the absence of the TO mode at ∼1489 cm$_{-1}$ in the xy configuration would also be very easy to detect.

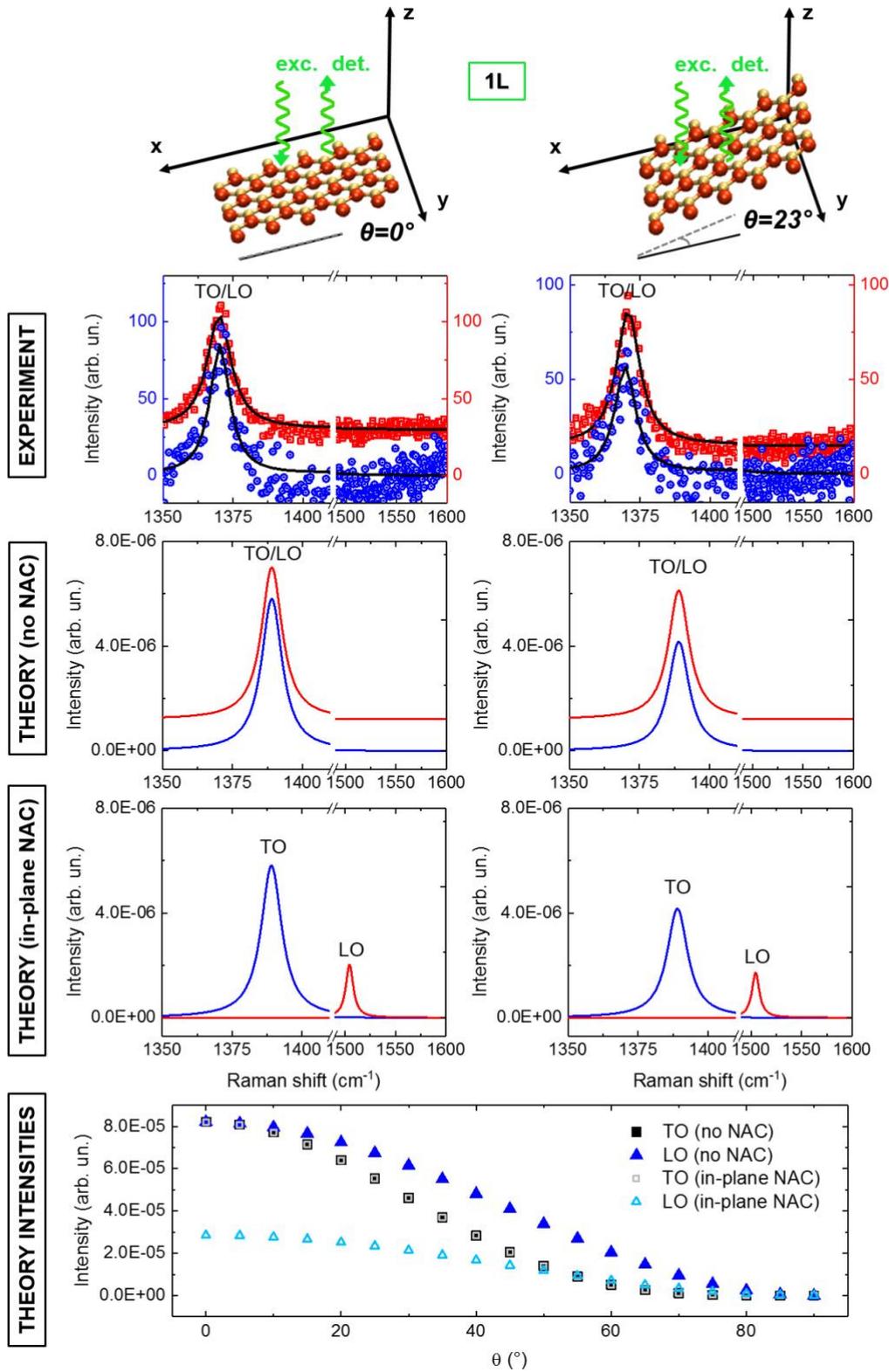

**Figure 4.** Experimental (first row) and calculated (second and third rows) Raman spectra of the 1L of h-BN in the xx (blue circles in the experiment, blue lines in the theory) and xy (red squares in the experiment, red lines in the theory) scattering geometries for planar flakes (left panels) or tilted flakes (right panels). The top insets sketch the experimental and theoretical scattering geometries, namely flakes either lying on the xy plane or forming an angle of 23° with the x axis. Red and yellow spheres represent B and N

atoms, respectively. The theoretical spectra are calculated without NACs (second row) and with NACs in the flake's plane (third row), imposing a FWHM of 9 cm$^{-1}$ for the single Lorentzian peaks. Experimental spectra were acquired with the 514 nm excitation wavelength. The increase in the background signal at 1600 cm$^{-1}$ for the xx spectra is due to the laser and indeed it is visible also when measuring on the substrate of the flakes. Black solid lines in the experimental spectra are fits to the data. In each plot, the scale for the xx and xy data is the same. An arbitrary offset has been applied to the intensity of the experimental spectra (and to theoretical spectra without NACs) in xy geometry for clarity purposes. The intensity scales in the theory spectra calculated without NACs spectra were arbitrarily chosen to resemble the experimental spectra for the planar flake and then the same scales were used for all the other calculated spectra. Notice the decrease in intensity in all the theory spectra induced by the titling of the flake. The bottommost panel shows the computed intensity of the TO (squares, in xx configuration) and LO (triangles, in xy configuration) modes in 1L h-BN as a function of the tilt angle with in-plane NACs (small, open symbols) and without NACs (big, filled symbols).

A clarification is now in order regarding the dependence of the intensity of the TO and LO modes on the tilt angle. Since NACs are applied in the flakes' plane, the frequency of the TO and LO modes and, therefore, the TO-LO splitting, does not depend on the tilt angle of the flake (see, *e.g.*, the third row in Figure 4). The intensity, on the other hand, depends on the angle, as it is shown more clearly in the bottommost panel in Figure 4, where we have plotted the calculated intensity of the TO and LO modes of the 1L as a function of the tilt angle, $\theta$. As anticipated in Figure 3 in the case of WSe$_2$, the change in the intensity is not a NAC-related effect. It rather derives from the different intensity of the modes in the different scattering geometries. Indeed, as the tilt angle increases, the intensity of the modes decreases, because in the limit of $\theta=90°$ the xx spectrum of the tilted sample must tend to the zz spectrum of the planar sample, where modes are either not allowed or have a negligible intensity. The different change in the intensity of the TO and LO modes without NACs is related to the fact that we tilt the flake about the y-axis. Therefore, in the case of the TO, which is seen in xx, tilting the sample we are rotating it with respect to both the incident and backscattered photons' polarization; in the case of the LO, which is seen in xy, the polarization of the backscattered photons does not change. Indeed, we have verified that if we detected the TO in the yy scattering geometry, which is equivalent to the xx, its intensity would be invariant with respect to the tilt angle chosen in our experiment.

For h-BN flakes containing two or more layers, two types of stacking along the z direction can be experimentally observed: AA' and AB, with AA' being the more frequent one [20]. They are both sketched in the central insets in Figure 5. As it is not straightforward to determine the type of stacking present in our flakes (as well as in the flakes investigated in the literature) given the small energy difference between the two types of stacking [20], we have to consider both types in the calculation of our Raman spectra. Figure 5 shows the experimental spectra (first row) obtained on a 3L h-BN flake at two different tilt angles (0° and 23°), both in the xx and xy scattering geometry. The calculated Raman spectra for the 3L with FWHM= 9 cm$_{-1}$ without NACs are displayed in the second row, and with NACs along the in-plane component of the photon **k**-vector in the third row. Dashed (solid) lines represent calculated spectra for the AB (AA') types of stacking. If the flake adopts an AB type of stacking, we obtain a situation similar to the 1L case: in the theory with NACs in xy geometry there is no mode visible at ∼1389 cm$_{-1}$, as the LO upshifts to ∼1578 cm$_{-1}$, which is in disagreement with the experimental observations and this would be a further proof of the necessity to avoid the application of NACs in the Raman spectra of 2D materials. If, conversely, the flake adopts an AA' type of stacking, in the spectra with in-plane NACs in xy geometry the TO mode does not disappear and the LO upshifts and has a negligible intensity (it is indeed almost invisible in our plots), both features being in apparent agreement with our experimental observations. Therefore, in case of AA' stacking, one could not establish whether the TO-LO spitting is suppressed or simply the LO mode shifts indeed to higher frequency, but is not detected by Raman scattering. We stress that the small downshift of the TO and LO modes in the AA' spectra compared to the AB spectra is due to the type of stacking and not to the application of NACs. The discussion relative to the decrease in the intensity due to the tilting of the flake carried out for the 1L sample applies also to the 3L sample.

We point out that if the number of layers were even, as in the 2L, the situation would be similar to the 3L case, the only difference being that upon application of NACs, in the xy geometry in case of AA' stacking, the LO would be Raman-forbidden rather than allowed with a small intensity (see Figure 4 in the SI5). The reason is that if the flake adopts an AA' stacking it has an inversion center, like in even-layer WSe$_2$, and just like in that case (depicted in Figure 2) the mode altered by an eventual NAC would be Raman inactive. Conversely, if the stacking is AB there is no inversion symmetry and the theoretical and experimental data can only be reconciled if the NACs are not included. Since it is not possible to reliably assess which type of stacking is adopted in our samples, the agreement between theory and experiment that we find when investigating the flakes thicker than a monolayer is not fully conclusive. Only the measurements displayed in Figure 4 for the 1L provide the unambiguous proof of the need to avoid application of NACs in h-BN and, consequently, of the absence of the TO-LO splitting.

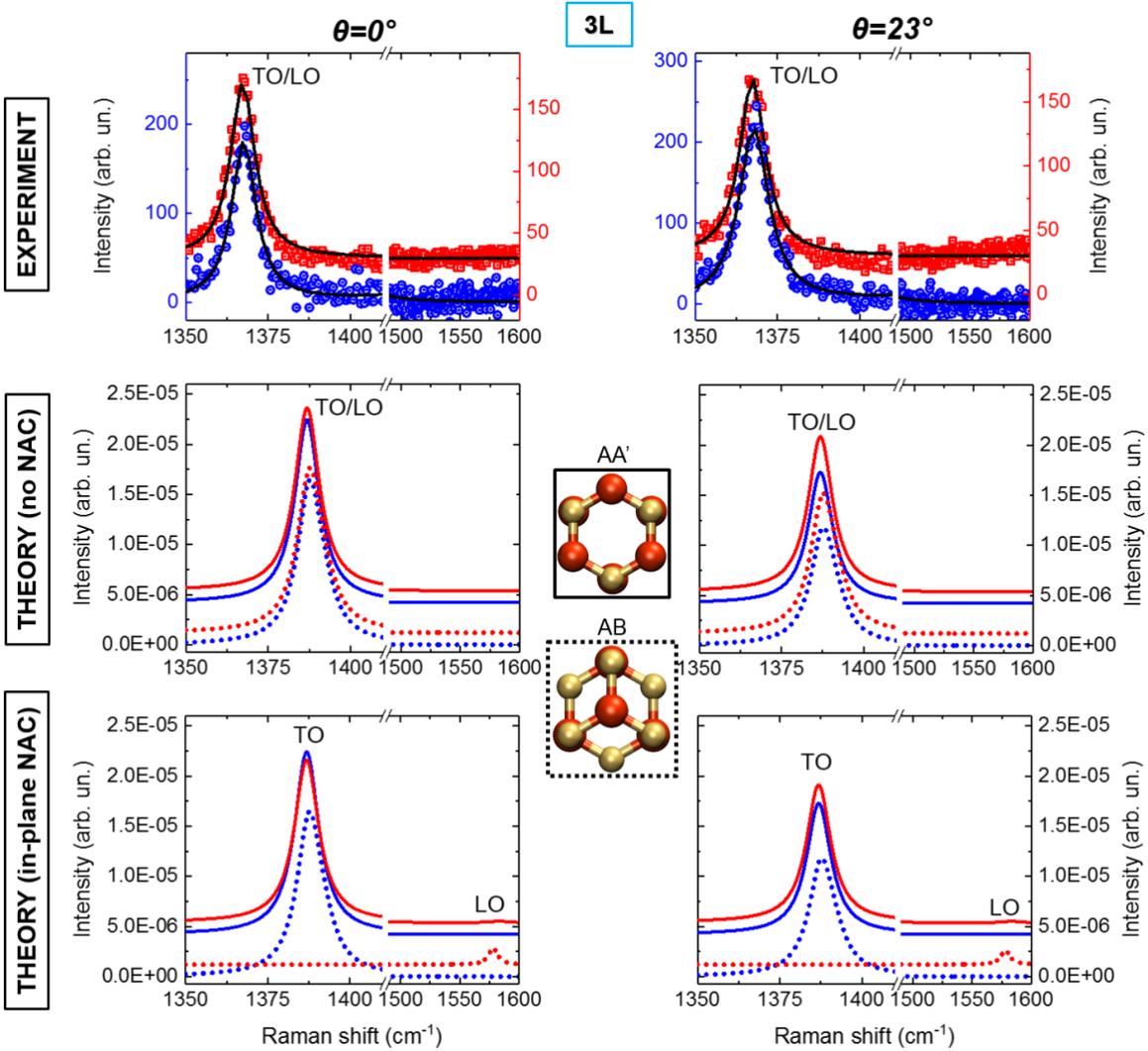

**Figure 5.** Experimental (first row) and calculated (second and third rows) Raman spectra of the 3L of h-BN in the xx (blue circles in the experiment, blue lines in the theory) and xy (red squares in the experiment and red lines in the theory) scattering geometries for planar flakes (left panels) and flakes tilted at 23° (right panels). The theoretical spectra are calculated without NACs (second row) and with NACs in the flake's plane (third row), imposing a FWHM of 9 cm$^{-1}$ for the single Lorentzian peaks. Spectra calculated with the AB (AA') stacking are represented by dashed (solid) lines. The central insets sketch the two types of stacking. Experimental spectra were acquired with the 514 nm wavelength. Black solid lines in the experimental spectra are fits to the data. In each theoretical plot, the scale for the xx and xy data is the same. An arbitrary offset has been applied to the intensity of the experimental and theoretical spectra in xy geometry for clarity purposes. The intensity scales in the theory spectra calculated without NACs were arbitrarily chosen to resemble the experimental spectra for the planar flake and then the same scales were used for all the other calculated spectra. An offset (the same for all the panels) to the theory spectra relative to the AA' stacking is introduced for clarity.

Finally, we attribute the lack of attention that the degeneracy of TO and LO modes has received in 2D materials so far to the serendipitous combination of circumstances (*i.e.* no splitting when the NACs are — incorrectly— applied along the *z* direction, little or zero Raman activity for the eventually split modes) that prevent a clear understanding of the Raman experiments.

## 3. Conclusions

We have presented a combination of Raman spectroscopy experiments and DFPT calculations of 1-4L $WSe_2$ and of 1L and 3L h-BN to experimentally verify the fundamentally different behavior in polar 2D *vs* 3D systems regarding the effect of the dipole−dipole interactions that translates into the need to avoid non-analytical corrections when properly calculating the Raman spectra of TMDs and polar 2D materials in general. As a peculiar result of inversion symmetry, the application of the NACs in the case of polar 2D materials with even number of layers might lead to results consistent with the experiments, but only because the affected modes would be Raman inactive. This latter effect might occur in certain materials (*e.g.* h-BN) also for odd number of layers depending on the type of stacking, because in the types of stacking providing the flakes with an inversion center (*e.g.* AA') the modes affected by NACs would have a negligible Raman intensity.

## 4. Methods

### *4.1 Preparation of samples and AFM characterization*

Our $WSe_2$ bulk crystals were purchased from hqgraphene. The samples were prepared by mechanical exfoliation followed by the dry-transfer of the flakes on a PDMS stamp [21]. Once the flakes were localized on the stamp by transmission measurements, they were transferred by using a micro-manipulator on Si with 90 nm of $SiO_2$ on top with micrometer resolution. In this way we have obtained the investigated 1L, 2L, 3L, and 4L flakes. The AFM characterization was performed in tapping mode using an AFM Neaspec set-up in ambient conditions. The processing and analysis of the experimental data was realized employing the WSxM software [22]. The h-BN flakes were obtained by exfoliating bulk BN crystals using the adhesive tape ELP BT-150P-LC supplied by Nitto. The flakes were transferred from the tape to a silicon wafer with an oxide thickness of 85nm, which was optimized for the optical contrast of monolayer h-BN. The thickness of the flakes was determined by AFM (Bruker Dimension 3100) measurements at ambient conditions and using tapping mode after transferring them to an atomically flat h-BN substrate. The results of the measurement are shown in Supplementary Figure 1. The used transfer technique is described in the SI.

All the samples are considerably larger than the diffraction-limited laser spot and far enough from samples having different layers' thickness, which made the spectroscopic measurements reliable even in the tilted geometry.

*4.2 Raman experimental details*

Raman measurements were performed in backscattering geometry on $WSe_2$ and h-BN flakes at room temperature, with excitation wavelength of 632.8 nm provided by a HeNe laser and of 514 nm provided by a $Ar_+Kr_+$ laser. The power was 0.1 mW for $WSe_2$ and 1.5 mW for h-BN (selected after checking that no heating/damaging effects were induced). The laser beam was focused with a 100x objective (numerical aperture of 0.80) and the measured spot size is diffraction-limited (< 1 μm). The scattered light was collected by a T64000 (Horiba) triple spectrometer in subtractive mode and equipped with 1800 g/mm gratings and liquid-nitrogen cooled multichannel charge couple device detector. Polarization-resolved Raman scattering experiments were performed with the incident and scattered light polarization vectors, $\varepsilon_i$ and $\varepsilon_s$, in the xy plane (x=100, y=010). Spectra were collected (and calculated) by selecting scattered light polarized either parallel or perpendicular to the polarization direction (x) of incident light, in the text indicated as xx and xy for simplicity. The xy plane is the plane of the sample when the sample is planar. For realizing the measurements on the tilted flakes, the flakes are mounted on a rotational stage (able to provide rotations in the xz plane with a resolution of ±2°). In both the planar and tilted geometries, the samples are mounted on a xy micrometric stage (with a resolution of 50 nm) for optimal positioning on the desired layers. All the Raman spectra were acquired with a 120s acquisition time for $WSe_2$ and 210 s for h-BN, and they were averaged from 3 to 5 times (depending on the noise level).

*4.3 Theoretical calculations*

We have calculated the ground-state geometry, the electronic structure and the Raman spectra of thin layers and bulk WSe2 and h-BN. In the case of WSe2 multilayers, we have considered the 2H structure, which is the most stable polytype. We have used the ABINIT code [23] to optimize the lattice vectors and the atomic position using norm-conserving pseudopotentials [24], a plane-wave cutoff of 41 and 50 Ha for WSe2 and h-BN, respectively, and a 16x16x1 **k**-point grid. We have used DFPT to compute the phonon dispersion relations and the Raman susceptibilities. For the bulk we have used 4 **k**-points along kz. The exchange-correlation energy has been calculated within the local density approximation (LDA) in the Ceperley-Alder parameterization. No van der Waals corrections have been included. It has previously argued that this computational setup yields frequencies in better agreement with experimental data [25].

Frequencies estimated by DFPT are at 0 K, while all the experimental ones are measured at room temperature. Considering that in WSe2 the A'1/E' modes upshift only by ~1 cm-1 [26] when going from room to low temperature, we have a nearly perfect agreement between theory and experiment; in h-BN the TO/LO modes upshift by ~5 cm-1 going from room to low temperature [27] and we have ~20 cm-1 shift between theory and experiment, similarly to previous LDA results [11]. This is not a serious limitation of the predictive nature of the theory results, as we have acquired the Raman spectra of h-BN in a very broad spectral region to make sure that the phonon modes arising from the application of NACs, if any, would have been clearly detectable.

In order to perform the calculations in the very same geometry of the experiments, the Raman intensity of each mode $n$ has been calculated as

$$I_n \propto |\varepsilon_i \, R_n \, \varepsilon_s|^2$$

where $R_n$ is the Raman susceptibility tensor calculated *ab initio* in the reference system of the experiment, while $\varepsilon_i$ and $\varepsilon_s$ are the polarization vectors of the incident and scattered light, respectively. This method is thoroughly explained in references [10][28]. The discussion on the application of NACs can be found in the main text. Once the intensity for each phonon mode has been calculated, Raman spectra are generated by summing up Lorentzian functions, each associated to a calculated mode frequency. The full width at half maximum used for the Lorentzian functions was 1.5 cm$^{-1}$ for WSe$_2$ and 9 cm$^{-1}$ for h-BN, chosen to resemble the experimental broadening. The intensities in the experimental spectra are directly comparable with each other, and so do the theoretical spectra with each other; however, experimental and theoretical intensities cannot be compared because different 'arbitrary units' are employed. Moreover, the relative intensity of spectra taken on samples with different number of layers is not meaningful because different number of layers result in different resonant Raman conditions involving excitonic effects [29][30], which are not taken into account in our calculations.

## Supporting Information

Details on the exfoliation of h-BN flakes and on their AFM characterization, calculated frequencies of phonon modes of WSe$_2$ 1L-4L samples, calculated Raman spectra with small FWHM of WSe$_2$ and h-BN with and without NACs. Effect of the type of stacking on the calculated Raman spectra on 2L and 3L h-BN samples.

**Acknowledgements**

This project has received funding from the Swiss National Science Foundation research grant (Project Grant No. 200021_165784). M. D. L. acknowledges support from the Swiss National Science Foundation Ambizione grant (Grant No. PZ00P2_179801). R.R. acknowledges financial support by the Ministerio de Economía, Industria y Competitividad (MINECO) under grant FEDER-MAT2017-90024-P and the Severo Ochoa Centres of Excellence Program under grant SEV-2015-0496 and by the Generalitat de Catalunya under grant no. 2017 SGR 1506. X.C. acknowledges financial support by the Ministerio de Economía, Industria y Competitividad under grant TEC2015-67462-C2-1-R (MINECO/FEDER), the Ministerio de Ciencia, Innovación y Universidades under Grant No. RTI2018-097876-B-C21 (MCIU/AEI/FEDER, UE), and the EU Horizon2020 research and innovation program under grant No. GrapheneCore2 785219. CS and DI acknowledge support by the Swiss Nanoscience Institute (SNI), the ERC project TopSupra (787414), the European Union Horizon 2020 research and innovation program under grant agreement No. 785219 (Graphene Flagship), the Swiss National Science Foundation and the Swiss NCCR QSIT. J.M.-S. acknowledges support through the Clarín Programme from the Government of the Principality of Asturias and a Marie Curie-COFUND European grant (PA-18-ACB17-29). R.T. Acknowledges support by European Union's Horizon 2020 research and innovation programme (SPQRel grant agreement no. 679183). K.W. and T.T. acknowledge support from the Elemental Strategy Initiative conducted by the MEXT, Japan and the CREST(JPMJCR15F3), JST.**Author Contributions**

M.D.L., X.C., and I.Z. conceived the experiment. J.M. and R.T. prepared and characterized by AFM the WSe$_2$ flakes. K. W. and T. T provided the h-BN samples, from which D.I. and C.S. prepared and characterized by AFM the flakes. M.D.L. performed the Raman measurements and analyzed the results. X.C. and R.R. performed the theoretical calculations. R.R., M.D.L., and I.Z. wrote the manuscript with contributions from all authors. All authors have given approval to the final version of the manuscript. ‡These authors contributed equally.

**References**

[1]  L. Du et al., *Strongly enhanced exciton-phonon coupling in two-dimensional WSe$_2$*, Phys. Rev. B **97**, 235145 (2018)

[2]  U. Wurstbauer, B. Miller, E. Parzinger, and A. W. Holleitner, *Light–matter interaction in transition metal dichalcogenides and their heterostructures*, J. Phys. D: Appl. Phys. 50 173001 (2017)

[3]  L. M. Malard, M. A. Pimenta, G. Dresselhaus, and M. S. Dresselhaus, *Raman spectroscopy in graphene*, Phys. Rep. **473**, 51-87 (2009)

[4]  S. Chen, Q. Wu, C. Mishra, J. Kang, H. Zhang, K. Cho, W. Cai, A. A. Balandin, and R. S. Ruoff, *Thermal conductivity of isotopically modified graphene*, Nature Mat. **11**, 203-207 (2012)

[5]  S. Baroni, S. de Gironcoli, A. Dal Corso, and P. Giannozzi, *Phonons and related crystal properties from density-functional perturbation theory*, Rev. Mod. Phys. **73**, 515 (2001)

[6]  G. Petretto, S. Dwaraknath, H. P. C. Miranda, D. Winston, M. Giantomassi, M. J. van Setten, X. Gonze, K. A. Persson, G. Hautier, and G.-M. Rignanese, *High-throughput density-functional perturbation theory phonons for inorganic materials*, Scientific Data **5**, 180065 (2018)21

# Supporting Information

# Experimental demonstration of the suppression of optical phonon splitting in 2D materials by Raman spectroscopy

Marta De Luca[1], Xavier Cartoixà[2], David I. Indolese[1], Javier Martín-Sánchez[3,4], Kenji Watanabe[5], Takashi Taniguchi[5], Christian Schönenberger[1,6], Rinaldo Trotta[7], Riccardo Rurali[8*], and Ilaria Zardo[1*]

[1] Departement Physik, Universität Basel, 4056 Basel, Switzerland
[2] Departament d'Enginyeria Electrònica, Universitat Autònoma de Barcelona, 08193 Bellaterra, Barcelona, Spain
[3] Department of Physics, University of Oviedo, 33007 Oviedo, Spain
[4] Center of Research on Nanomaterials and Nanotechnology, CINN (CSIC−University of Oviedo), El Entrego 33940, Spain
[5] National Institute for Material Science, 1-1 Namiki, Tsukuba 305-0044, Japan
[6] Swiss Nanoscience Institute, University of Basel, 4056 Basel, Switzerland
[7] Department of Physics, Sapienza University of Rome, 00185 Rome, Italy
[8] Institut de Ciència de Materials de Barcelona (ICMAB–CSIC), Campus de Bellaterra, 08193 Bellaterra, Barcelona, Spain

*Authors for correspondence, E-mail: rrurali@icmab.es; ilaria.zardo@unibas.ch

1. Exfoliation and characterization of h-BN flakes

2. Calculated phonon frequencies and intensities of WSe2 at $\Gamma$ point

3. Calculated Raman spectra of WSe2 in the region of the ~250 cm$_{-1}$ phonon modes

4. Sketches of the atomic displacements of the phonon modes of 1L and 2L WSe2 in the region of the ~176 cm$_{-1}$ and ~310 cm$_{-1}$ phonon modes

5. Calculated Raman spectra of h-BN in the region of the ~1380 cm$_{-1}$ phonon modes

# 1. Exfoliation and characterization of h-BN flakes

Exfoliation

The hexagonal boron-nitride (h-BN) flakes are obtained by exfoliating bulk BN crystals using the adhesive tape ELP BT-150P-LC supplied by Nitto. The transfer from the tape to the Si/SiO$_2$ substrate was done by a technique established by refs. [1][2] using a polycarbonate (PC) film on a polydimethylsiloxane (PDMS) pillow. For AFM characterization, the flake was picked up from the SiO$_2$ at 80°C and released on an atomically flat h-BN substrate by heating the PDMS/PC stamp above 160°C. Above this temperature the PC film starts to melt and detaches from the PDMS pillow. The h-BN flake covered with PC remains on the new substrate. The PC was dissolved in dichloromethane for 1 hour. Afterwards the sample was cleaned in isopropanol and blow dried with nitrogen. In a next step the sample was annealed in an Ar/O$_2$ (90%/10%) atmosphere at 500°C and a background pressure of 100mBar for 3 hours to clean its surface from remaining polymer residues [3].

Characterization

The thickness of the flakes was determined by AFM, as shown in Supplementary Figure 1. For this measurement, the flakes were transferred from the silicon oxide to an atomically flat h-BN substrate after the Raman measurements, which was exfoliated as described above. This procedure has the advantage that the flake is lying flat on the substrate and no elevation due to water or a contamination layer is expected [4]. Furthermore, there is no difference in the tip-surface interactions between the substrate and the measured material. With this analysis, we have identified the monolayer and trilayer sample that have been investigated by Raman spectroscopy (see the images corresponding to the red and the blue squares, respectively).

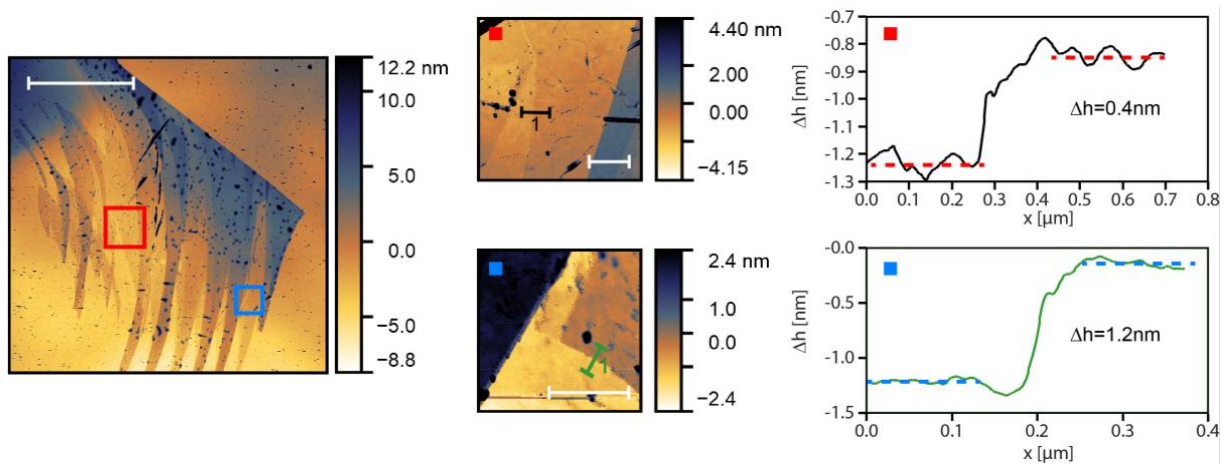

**Supplementary Figure 1.** Left panel: AFM scan of the whole h-BN flake after exfoliating a BN crystal on SiO$_2$ substrate. Scale bar is 10 μm. Central panels: AFM scans of the regions marked with red and blue squares after transferring the h-BN flake to a BN substrate. The bottom part of the flake in the blue square was ripped off during the transfer to BN, resulting in a straight edge. Scale bar is 1 μm. Right panels: Line cuts of AFM measurements shown in the central panels, taken across the regions marked by '1' bars. The thickness of a monolayer is known to be ~0.4 nm [4], therefore, our profiles correspond to a 1L (red square) and a 3L (blue square). The dashed lines are indicating the levels which were used for extracting the height difference.

## 2. Calculated phonon frequencies and intensities of WSe$_2$ at Γ point

Supplementary table 1 shows the frequencies of the phonon modes at Γ point of the 1L, 2L, 3L, and 4L samples calculated by density functional theory (DFPT) without non-analytical corrections (NAC). The total number of branches is 9 for the 1L, 18 for the 2L, 27 for the 3L, and 36 for the 4L, as expected based on the number of atoms per unit cell (3, 6, 9, and 12, respectively). For clarity, the corresponding branch labels are given only for the 1L in parentheses. The Raman active modes allowed in our xx and xy scattering geometries are highlighted in yellow, as detailed in Supporting Information 3.

| 1L | 2L | 3L | 4L |
|---|---|---|---|
| 0 (ZA) | 0 | 0 | 0 |
| 0 (TA) | 0 | 0 | 0 |
| 0 (LA) | 0 | 0 | 0 |
| 176.17 (TO$_1$) | 17.74 | 12.39 | 9.43 |
| 176.17 (LO$_1$) | 17.74 | 12.39 | 9.43 |
| 250.76 (TO$_2$) | 27.34 | 19.26 | 14.73 |
| 250.76 (LO$_2$) | 175.54 | 21.63 | 17.59 |
| 251.31 (ZO$_2$) | 175.54 | 21.63 | 17.59 |
| 312.32 (ZO$_1$) | 176.43 | 33.21 | 23.03 |
|  | 176.43 | 175.27 | 23.03 |
|  | 249.57 | 175.27 | 27.17 |
|  | 249.57 | 175.97 | 35.34 |
|  | 249.68 | 175.97 | 175.14 |
|  | 249.68 | 176.51 | 175.14 |
|  | 250.41 | 176.51 | 175.64 |
|  | 251.79 | 248.57 | 175.64 |
|  | 310.50 | 248.57 | 176.19 |
| | 311.04 | 249.60 | 176.19 |
| | | 249.60 | 176.54 |
| | | 249.61 | 176.54 |
| | | 249.61 | 248.49 |
| | | 250.08 | 248.49 |
| | | 251.07 | 248.61 |
| | | 252.01 | 248.61 |
| | | 309.19 | 249.59 |
| | | 310.71 | 249.59 |
| | | 310.81 | 249.59 |
| | | | 249.59 |
| | | | 249.91 |
| | | | 250.62 |
| | | | 251.47 |
| | | | 252.12 |
| | | | 308.89 |
| | | | 309.62 |
| | | | 310.67 |
| | | | 310.81 |

**Supplementary Table 1:** Frequencies (in cm-1) of the phonon modes calculated at the Γ point of the WSe2 monolayer, bilayer, trilayer, and quadrilayer by DFPT. All the modes that are Raman-allowed in the xx or the xy scattering geometry are highlighted in yellow.

### 3. Calculated Raman spectra of WSe2 in the region of the ~250 cm-1 phonon modes

In Figure 1 in the main text we display Raman spectra in the region of the ~250 cm-1 phonon modes of 1L, 2L, 3L, 4L, and bulk of WSe2 calculated by DFPT without NAC and obtained by setting a full width at half maximum (FWHM) of the Lorentzian peaks equal to 1.5 cm-1, in order to reproduce the experimental broadening (as explained in the methods section). However, as displayed in Supplementary Table 1, there are several Raman active phonon modes contributing ―each one with a certain intensity that we have calculated― to the Raman signal in that spectral region, and the spectral broadening that we impose usually hides the single contributions. Therefore, in Supplementary Figure 2, we display the calculated Raman spectra in the xx and xy scattering geometries with FWHM of 0.1 cm-1 (left panel) and with FWHM of 1.5 cm-1 as in the main text (right panel), thus clarifying which modes contribute to the broadened Raman spectra displayed in Figure 1 in the main text.

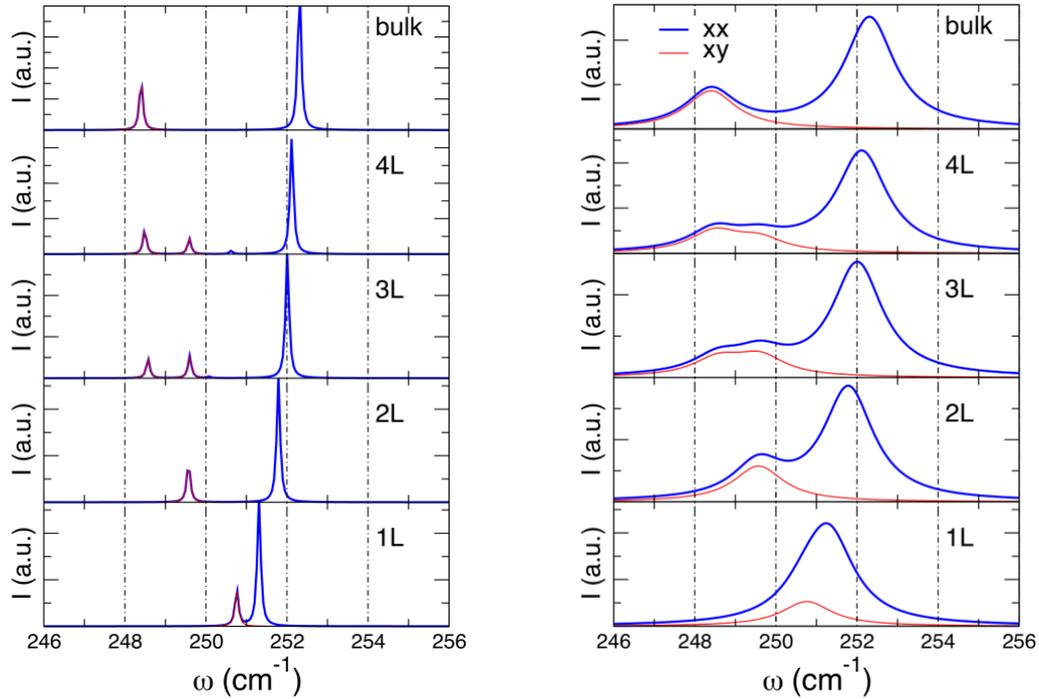

**Supplementary Figure 2.** Calculated Raman spectra of 1L, 2L, 3L, 4L, and bulk of WSe$_2$ obtained by setting the FWHM of the single Lorentzian peaks to 0.1 cm$^{-1}$ in the left panel and to 1.5 cm$^{-1}$ in the right panel (blue spectra: xx geometry; red spectra: xy geometry). Spectra were calculated without NACs.

## 4. Sketches of the atomic displacements of the phonon modes of 1L and 2L WSe$_2$ in the region of the ~176 cm$^{-1}$ and ~310 cm$^{-1}$ phonon modes

Figure 1 in the main text shows that the application of NAC has a remarkable effect on the phonon modes at ~250 cm$^{-1}$ (the reason why this is the case only in samples with an odd-number of layers is discussed in Figure 2 in the main text). Supplementary Figure 3 explains why in WSe$_2$ the application of NAC does not have any effect on the Raman spectra in the region of the ~176 cm$^{-1}$ and ~310 cm$^{-1}$ phonon modes, regardless on the parity of the number of layers: in the 1L there are no allowed modes in our geometries in those spectral regions and anyhow the application of in-plane NAC does not affect them; in the 2L there are allowed modes (those whose displacements are contained in yellow squares) but the NAC does not shift them; in the 3L and 4L a behavior similar to the 2L occurs.

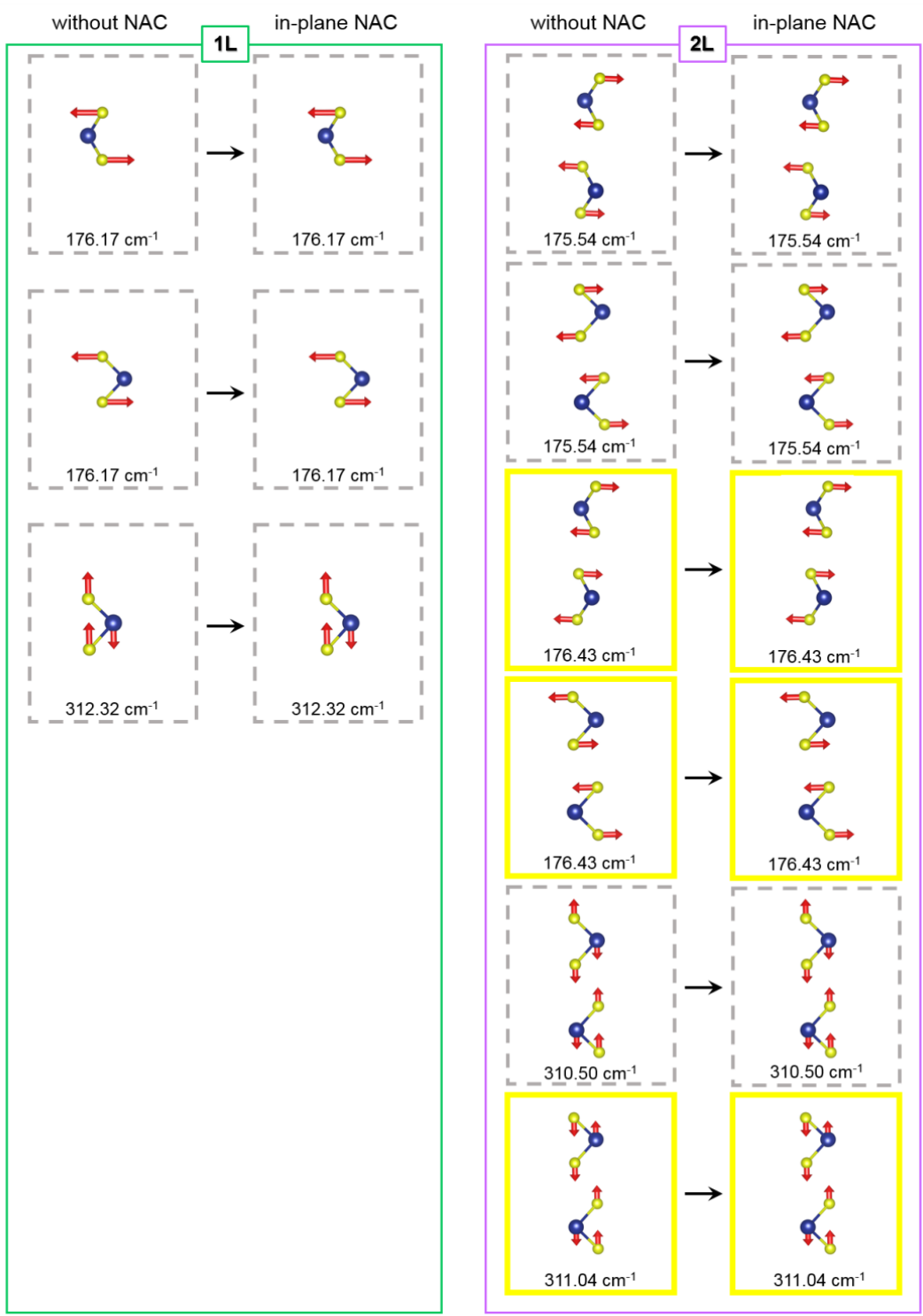

**Supplementary Figure 3.** Sketch of the atomic displacements of all the phonon modes but those at already shown in Figure 2 in the main text for the 1L (left panel) and 2L (right panel) samples, without NAC and with NAC in the flakes' plane. Displacements contained in yellow, continuous frames correspond to Raman-active modes allowed in our xx and xy scattering geometries, while displacements contained in grey, dashed frames correspond to modes that are either Raman-inactive or active but forbidden in our geometries and thus not showing up in the spectra in Supplementary Figure 2. Blue and yellow spheres represent W and Se atoms,

respectively. Black arrows are a guide to follow the effect of the application of in-plane NAC on each mode.

## 5. Calculated Raman spectra of h-BN in the region of the ~1380 cm$^{-1}$ phonon modes

In Figure 4 in the main text we display calculated Raman spectra of the h-BN measured samples, namely 1L and 3L, obtained by setting a FWHM in the Lorentzian peaks equal to 9 cm$^{-1}$, in order to reproduce the experimental FWHM. The imposed broadening hides the presence of low-intensity peaks that show up very close in frequency to the most intense ones. Therefore, in Supplementary Figure 4, we display the calculated Raman spectra in the xx and xy scattering geometries with FWHM of 0.1 cm$^{-1}$ for 1L, 2L, and 3L planar flakes, without NAC (left panel) and with in-plane NAC (right panel), thus clarifying which modes contribute to the broadened Raman spectra along with the effect of in-plane NAC on each mode.

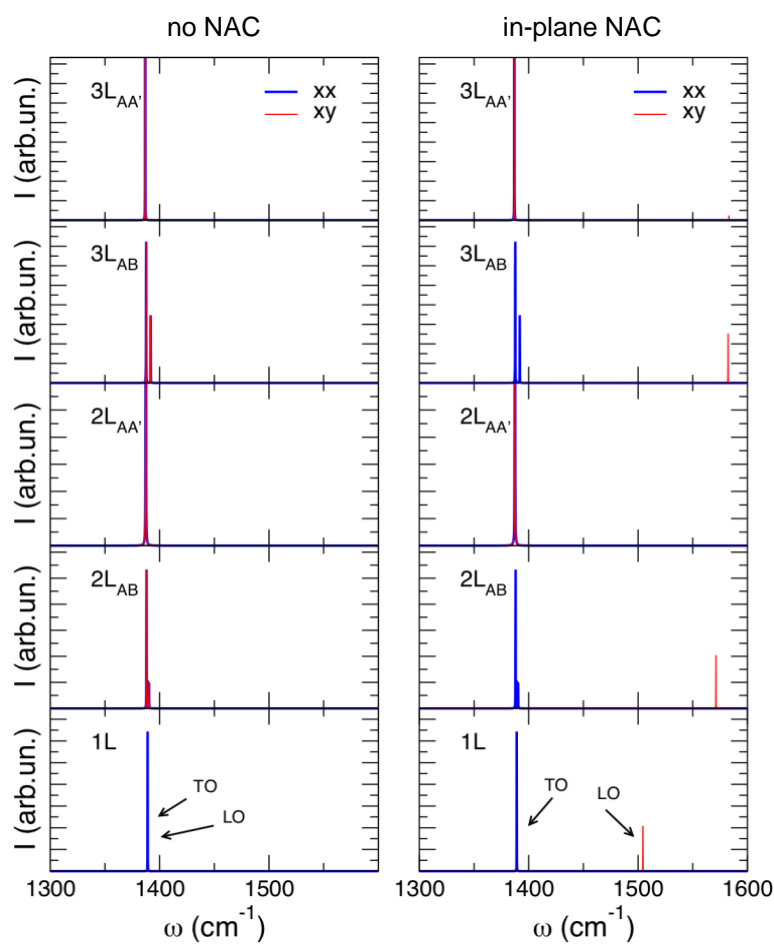

**Supplementary Figure 4.** Calculated Raman spectra of 1L, 2L, and 3L of h-BN obtained by setting the FWHM of the single Lorentzian peaks to 0.1 cm$^{-1}$ (blue spectra: xx geometry; red spectra: xy geometry). Spectra were calculated without NAC (left panel) and with NAC (right panel) applied in the flake's plane. The flake is considered to be flat on the xy plane.